# Developing the Field Selector System for the SHARP Multi-IFU: Optomechanical Design and First Prototype Validation

**H. Mahmoodzadeh[1,2*], P. Saracco[1], I. Di Antonio[3], P. Conconi[1], D. Scaccabarozzi[2], B. Saggin[4], E. Molinari[1], C. Arcidiacono[5], E. Cascone[6], V. Cianniello[6], F. D'Alessio[7], V. De Caprio[6], G. Di Rico[3], B. Di Francesco[3], C. Eredia[6], P. Franzetti[8], M. Fumana[8], E. Portaluri[3], F. Vitali[7],**

[1]*INAF - Osservatorio Astronomico di Brera, Via Brera 28, Milano, Italy, 20121*

[2]*Politecnico di Milano, Department of Mechanical Engineering, Via la Masa 1, Milano, Italy, 20156*

[3]*INAF - Osservatorio Astronomico d'Abruzzo, Via Mentore Maggini, Teramo, Italy, 64100*

[4]*Università degli Studi di Padova Department of Industrial Engineering Padova Italy*

[5]*INAF - Osservatorio Astronomico di Padova, Vicolo dell'Osservatorio 5, Padova, Italy, 35122*

[6]*INAF - Osservatorio Astronomico di Capodimonte, Salita Moiariello 16, Napoli, Italy, 80131*

[7]*INAF - Osservatorio Astronomico di Roma, Via Frascati, 33, Monte Porzio Catone (RM), Italy, 00040*

[8]*INAF - IASF, Via Alfonso Corti 12, Milano, Italy, 20133*

*Hossein.Mahmoodzadeh@Polimi.it

**This paper presents the innovative Field Selector System (FSS) developed for VESPER, the multi-Integral Field Unit of the SHARP near-infrared spectrograph. The proposed system employs an array of movable light-collection probes (Field Selectors) that travel along parallel, vertically offset planes to avoid mechanical interference and vignetting. The true novelty lies in its opto-mechanical synchronization: as each probe translates by $\Delta y$ to select a target, a set of folding mirrors shifts by $\Delta y/2$. This maintains a constant optical path length, allowing multiple independently moving probes to share a single focal plane and form a continuous image on it, easy to slice by the slicer. A dedicated optical test setup was developed to characterize the positioning stability and repeatability of the probe motion. The measured mechanical errors were converted into equivalent optical tilts and incorporated into the SHARP optical model to assess their impact on instrument performance. The experimental results identified the dominant sources of positioning error and provided valuable insight into the relationship between mechanical tolerances and optical performance. The study also highlights several design improvements capable of significantly reducing probe tilting and enhancing system stability. The results demonstrate the viability of the proposed concept and indicate that, with the identified improvements, the system represents an excellent field-selection solution for SHARP and other multi-integral-field-unit (mIFU) spectrographs.**

## 1. INTRODUCTION

The SHARP [1,2] near-infrared (NIR) multi-mode spectrograph is a cutting-edge instrument under development to meet the demanding requirements of next-generation astronomical observations. SHARP has been primarily conceived for the European Southern Observatory's Extremely Large Telescope (ELT), but its design is versatile and can be potentially integrated in any next-generation telescope that provides an adaptive optics correction across a large field of view.

SHARP comprises two primary components: NEXUS, a Multi-Object Spectrograph (MOS), and VESPER, a multi-Integral Field Unit (mIFU). This paper focuses on the development of the Field Selector System (FSS), one of the most critical sub-systems within VESPER. It is an innovative system designed to simultaneously collect light from multiple targets in the ELT focal plane and direct it to an image slicer. Image slicing is a key technology in modern astronomical instrumentation. By dividing a two-dimensional field of view into narrow, contiguous slices, an image slicer enables the efficient packing of spectra onto a detector, thereby maximizing spatial resolution, sensitivity, and throughput. This technique has been successfully implemented in several instruments, such as MUSE [3,4], GNIRS IFU [5], and NIRSpec [6], although these instruments employ substantially different field-selection and optical architectures.

In particular, the field-selection and light-coupling approaches used in existing IFU and multi-object spectrographs can be broadly classified into fixed field-splitting, discrete aperture selection, field scanning, and mechanically configurable pick-up systems. MUSE [3,4], for example, uses a fixed field-splitting and relay-optics architecture in which the input field is divided into 24 optical channels, with each channel subsequently reformatted by an image slicer. Similar image-slicer-based IFU approaches have also been implemented in instruments such as the GNIRS IFU [5] and Gemini instruments [7]. Other large-field spectrograph concepts, such as the Wide-Field Optical Spectrograph (WFOS) [8] for the Thirty Meter Telescope (TMT) [9], also employ image-slicing concepts, although their optical configurations and scientific requirements differ from those of SHARP. These approaches provide highly efficient spatial reformatting, but they do not employ movable light-collection probes that scan the focal plane and sequentially couple selected fields to a common IFU channel.

Field-scanning architectures provide a different approach to integral-field spectroscopy. The GRIS spectrograph [10,11] at GREGOR [12] is designed to scan a large field of view (FOV) in both vertical and horizontal directions. It utilizes an entrance window of 6.75 arcsec × 3 arcsec to scan a 1.0 × 1.0 arcmin FOV, sending the collected beam to the reimaging subsystem of the IFU optical bench. Similarly, the High-Resolution Integral Field Spectrograph (HRIFS) [13], installed on the European Solar Telescope (EST) [14] employs a 2D field scanning system with an 80 arcsec² entrance area, covering a 2 × 2 arcmin² FOV for detailed spectroscopic observations. These instruments demonstrate the capability of scanning systems to extend the effective field of view beyond the instantaneous entrance aperture. However, their optical and mechanical implementations differ from the probe-based light-collection concept proposed for SHARP.

Another relevant approach is represented by KMOS [15], which employs 24 independently configurable pick-up arms. Each arm positions a pick-off mirror at a selected location in the focal plane and feeds the collected light to a dedicated IFU. This architecture provides flexible target selection over a relatively large patrol field, but requires a separate positioning mechanism and optical channel for each configurable arm.

The SHARP FSS combines aspects of these established approaches while adopting a different system architecture. Six movable light-collection probes are associated with each VESPER channel and scan the focal plane to select the required target positions. The selected light is then directed toward the corresponding IFU optical path. Unlike fixed field-splitting systems such as MUSE, the SHARP FSS does not simultaneously divide the complete focal plane into a fixed set of channels and target selection is achieved through macroscopic movable optical elements rather than individually addressable focal-plane apertures. Furthermore, although field scanning is employed in instruments such as GRIS and HRIFS, the SHARP FSS uses movable probes for direct free-space light collection and coupling to a common VESPER IFU channel.

These differences are summarized in Table 1, which compares representative state-of-the-art instruments in terms of field-selection principle, optical coupling approach, number of active selection elements, field coverage, and relevant positioning or scanning characteristics.

These advanced systems demonstrate the versatility of image-slicing, field-scanning, and configurable pick-up techniques for optimizing light collection and spatial information in modern spectrographs. They also highlight the different trade-offs between instantaneous field coverage, target-selection flexibility, mechanical complexity, and optical coupling efficiency.

**Table 1: Comparison of representative field-selection and integral-field spectrograph architectures with the proposed SHARP Field Selector System (FSS).**

| Instrumen | Architecture | Key characteristics |
|---|---|---|
| **SHARP FSS** | Multi-IFU/linear scanning system | VESPER comprises 2 modules summing up to 12 probes. They can be deploys over an area (patrol field) of 20"x40". |
| **KMOS** | Multi-IFU Configurable pick-up arms | 24 arms; multiple separated targets; each arm feeds a dedicated IFU; ~7.2′ patrol field |
| **MUSE** | Single-IFU Fixed field splitting | complete contiguous field; relay optics + image slicers; 1′ × 1′ FOV |
| **NIRSpec** | Single-IFU | One square hoe 3"x3" |
| **GRIS** | Single-IFU/2D field scanning | 6.75″ × 3″ instantaneous field; one contiguous field per scan position; ~60″ × 60″ scanned FOV |
| **HRIFS** | Single-IFU/2D field scanning | ~80 arcsec² instantaneous field; one contiguous field per scan position; 2′ × 2′ scanned FOV |

The scientific requirements for SHARP, detailed in Table 2 and discussed by Saracco et al. [1,2], drive the need for innovative solutions in both optical and mechanical design. In particular, the VESPER concept requires a field-selection mechanism capable of accurately positioning multiple probes within the focal plane while maintaining efficient light collection and compatibility with the overall instrument architecture. To address these requirements, this study presents and experimentally investigates a new probe-based approach for field selection and light collection.

This paper is structured as follows: Section 2 describes the overall design and functional principles of SHARP's Field Selector System. Section 3 presents the mechanical concept design. Section 4 details the prototype development, experimental setup, and test methodology. Section 5 presents experimental results, analyzes the identified error sources and their impact on optical performance, and discusses potential improvements for future system development.

**Table 2: SHARP main requirements resulting from the scientific drivers.**

| | |
|---|---|
| Wavelength limit | 2.45 µm |
| Spectral ranges (simultaneous) | 0.95-2.45 µm |
| Angular resolution | ~ 30 mas |
| AO-corrected FoV | ~ 1x1 arcmin |
| Observing modes | MOS + multi-IFU |
| Spectral resolution | R>1000 |

## 2. SYSTEM CONCEPT

The designated area on the ELT's focal plane for VESPER is approximately 20"x40". Light from the MCAO system MORFEO [16,17] is directed toward the light-collecting system. Figure 1 (left) depicts six probes feeding one module of VESPER, covering half of the total area (~10"×40"). Each probe, called Field Selector (FS), consists of a prism whose entrance surface is shaped to provide optical power (effectively acting as a lens), and a collimator which moves rigidly by Δy along the y-axis (Figure 1 center). Two 45-degree movable mirrors bend the beam and adjust by Δy/2 each, maintaining a constant optical path between the collimator and the camera as presented in formula 1, so that all probes share the same focal plane. On this focal plane, the six FSs form a contiguous image of approximately 1.5"×10" (Figure 1 right).

$$L = a + b + c = a + \left(b + \frac{\Delta y}{2}\right) + \left(c - \frac{\Delta y}{2}\right) \qquad (1)$$

Figure 2 displays the optical design of the field selector system (FSS) feeding one VESPER channel, consisting of six probes. These probes are arranged along parallel planes, alternately offset up and down by about 24 mm. Additionally, the prisms are paired and oriented in opposite directions to direct light effectively. This design minimizes interference and prevents vignetting between components. The specifications of the optical components used in the FSS are summarized in Table 3.

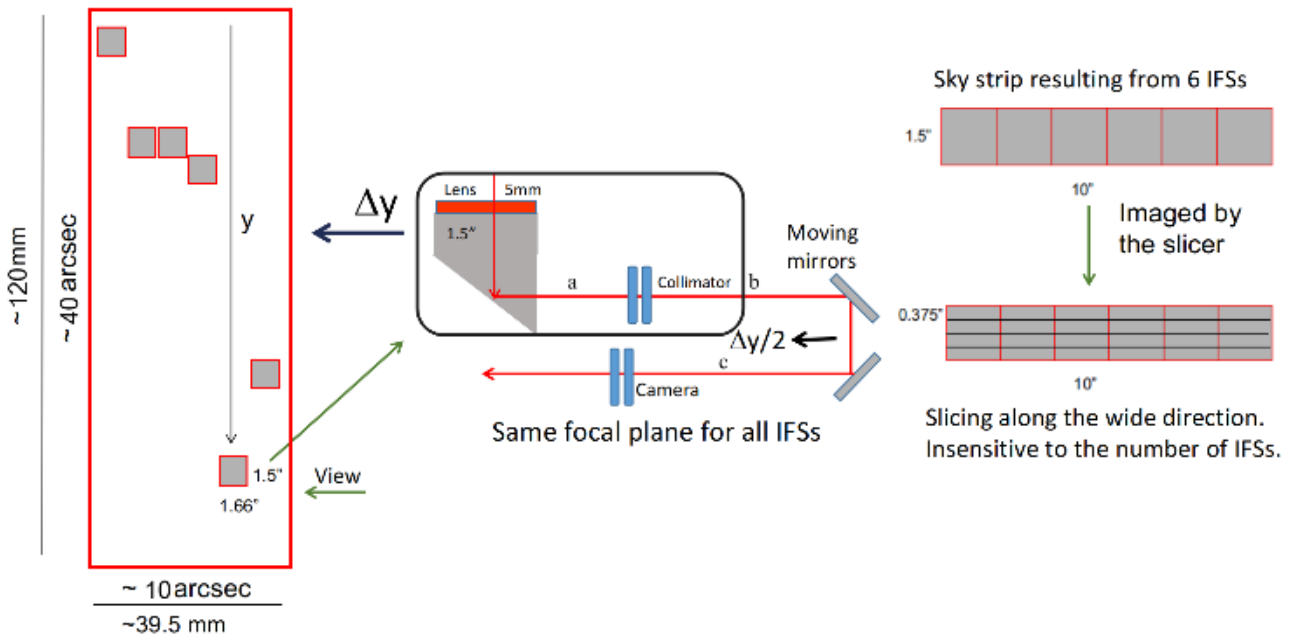

**Fig 1** Schematic view of the Field Selector System feeding one channel of VESPER. Left - The 6 FSs are aligned along the x direction. The FSs can be moved along the y direction to sample an area of ≈10"×40". Center - Each FS is composed of a prism having a spherical upper surface with power and a collimator. They rigidly move by Δy. Their movement is compensated by the movement (Δy/2) of the two 45-degree moving mirrors to keep constant the optical path between collimator and camera for all the FSs. Right – The image formed onto the VESPER focal plane by the 6 FSs is a strip as if they were contiguous.

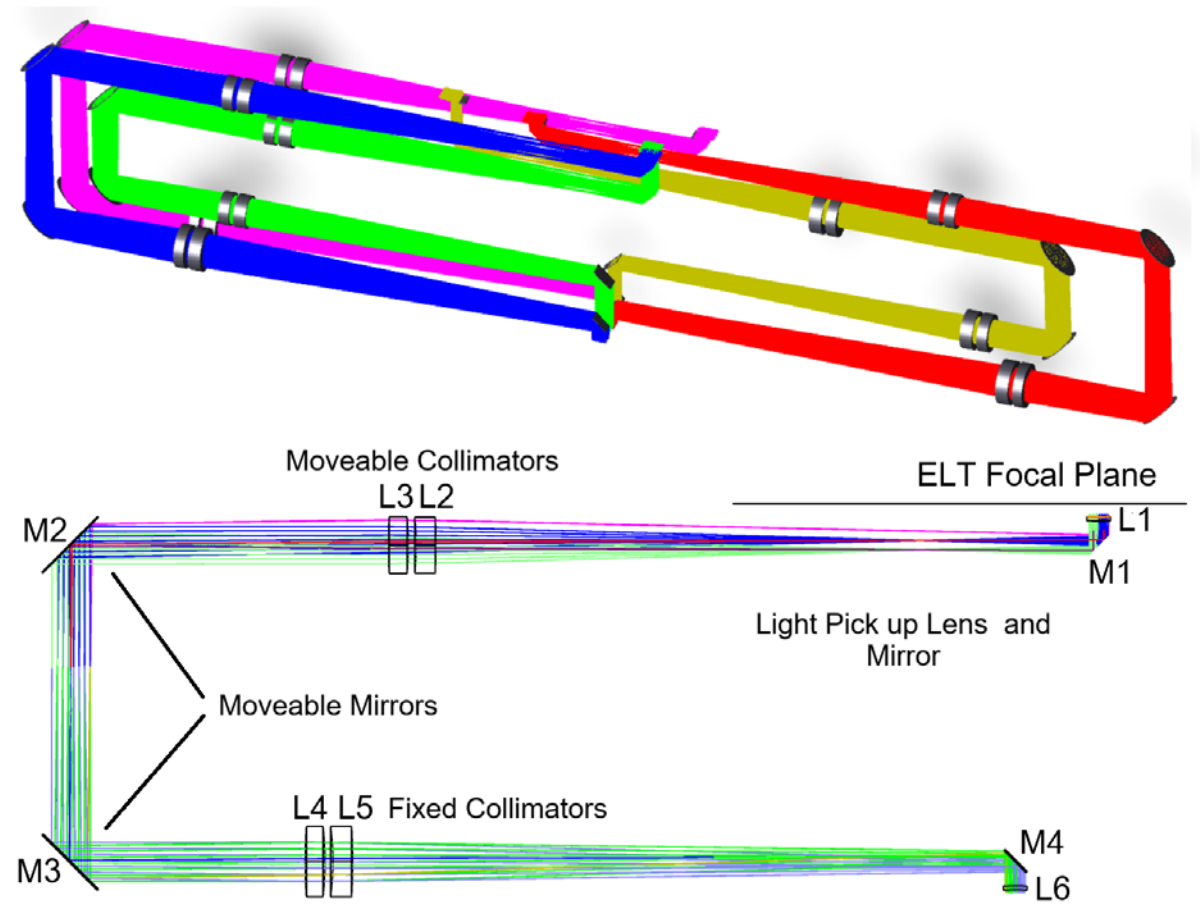

**Fig 2.** Optical path of the Field Selector System (FSS), where light is collected using six probes and delivered as a single long rectangular strip to the image slicer.

The light coming from the FSS of VESPER after an anamorphic relay is focused on the slicer, conceptually similar to the slicer of MUSE at VLT [3,4]. The slicer of VESPER is composed of four sets of 72 mirrors (the slicer of MUSE is made of sets of 48 mirrors). The mirror sizes are 1.0mm×22 mm. Figure 3 shows the design of two sets of mirrors. Each set of 72 mirrors samples a slice 0.375"×10" of the image along the wide direction. There is a corresponding set of pupil mirrors for each set of mirrors that converges the light toward the camera. The signal from each spaxel (31 mas) is fed into the spectrograph that generates a spectrum for each one of them. There is a camera feeding each set of 72 mirrors, therefore 4 cameras. The parameters reported here (10×40 arcsec F.o.V. and four sets of 72 mirrors of 1×22 mm) reflect the current consolidated opto-mechanical baseline of VESPER, superseding preliminary optical assessments, as also detailed in Saracco et al. [18].

**Table 3. Optical design parameters of the Field Selector System (FSS) for one VESPER channel.**

| | R1[mm] | R2[mm] | Thickness[mm] | Material |
|---|---|---|---|---|
| L1 | 26.28 | 23.69 | 1 | SILICA |
| L2 | 143.78 | 69.97 | 5 | SILICA |
| L3 | 79.84 | -125.68 | 5 | CAF2 |
| L4 | 125.68 | -79.84 | 5 | CAF2 |
| L5 | -69.97 | -143.78 | 5 | SILICA |
| L6 | -23.69 | 26.28 | 1 | SILICA |

## 3. CONCEPTUAL MECHANICAL DESIGN

The conceptual mechanical design of the FSS is depicted in Figure 4 and is crucial for the selection of targets and management of light within the optical system. This design employs an array of 12 prisms.

In Figure 5, the proposed mechanism for scanning the focal plane of VESPER is illustrated. For the first plane, the collecting prisms and their collimators are mounted on a rigid frame. This frame is connected to a sturdy guide rail that enables movement in the y direction, powered by a stepper motor and a lead screw mechanism, as shown in the figure.

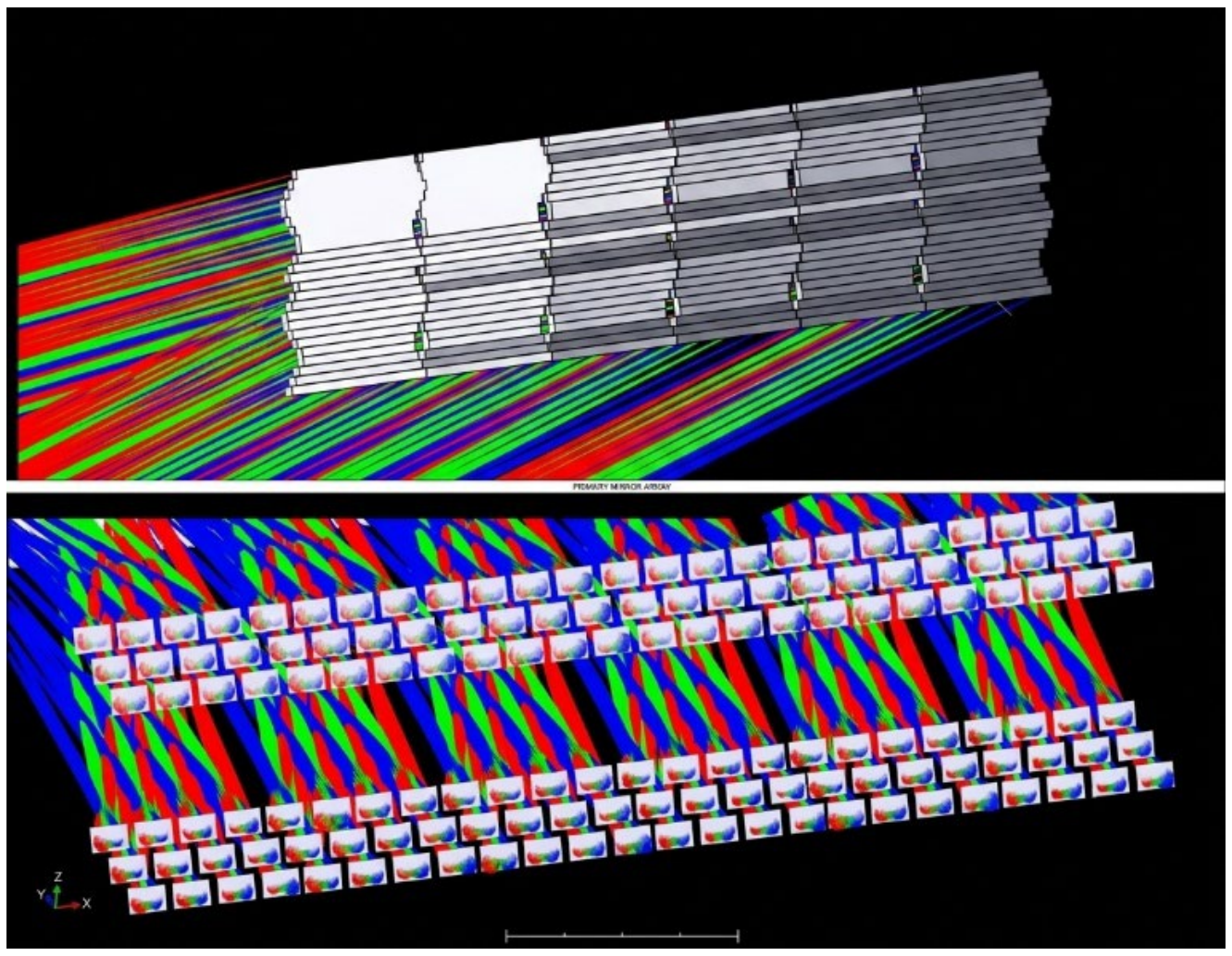

**Fig. 3** Slicer onto the focal plane of VESPER. The slicer is composed of four sets of 72 mirrors each. Two sets are shown on the top image, as example. Each set samples a slice of 0.375”×10” of the image provided by the 6 FSs. Particular of two sets of pupil mirrors as shown in bottom image, bringing the light to the corresponding cameras.

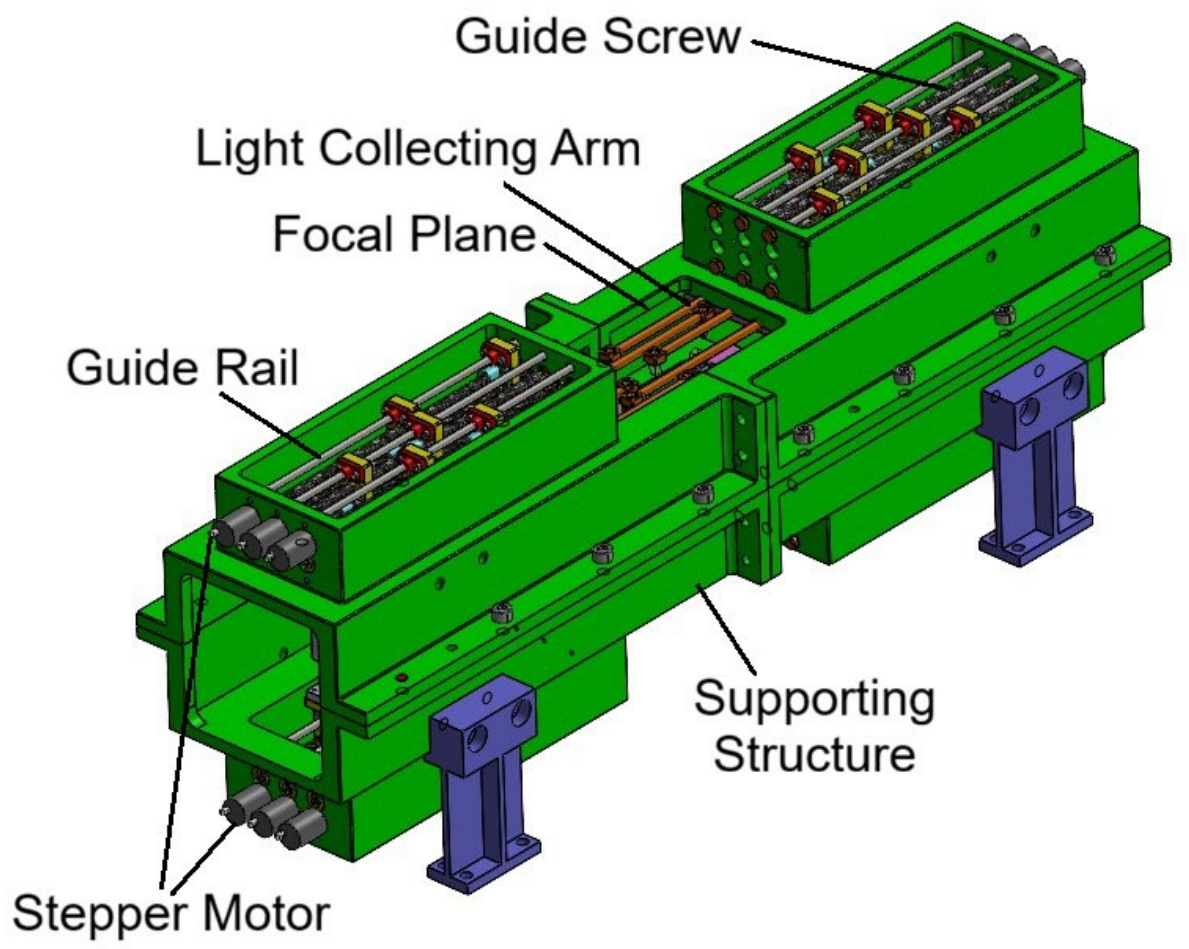

**Fig. 4** Conceptual mechanical design of the FSS, showing the arrangement of 12 selector prisms distributed over two parallel planes for target selection and light management within the optical system.

The movement of these folding mirrors is also driven by a ball screw mechanism. However, the same stepper motor that drives the collecting prism is utilized to move the folding mirrors, with the motion transmitted through a gear pair. Since the folding mirrors are required to move at half the displacement of the collecting prism, a gear ratio of 0.5 is implemented. Alternatively, the movement of the folding mirrors can be achieved using an independent stepper motor, similar to the one used for the collecting prism.

While the stepper motor is equipped with encoder to enable closed-loop position control, an additional location sensor could be installed on the guide rail for the folding mirrors to ensure precise control.

For the second plane, the same design concept is applied, with the key difference being that the guiding mechanism is installed at the bottom of the structure.

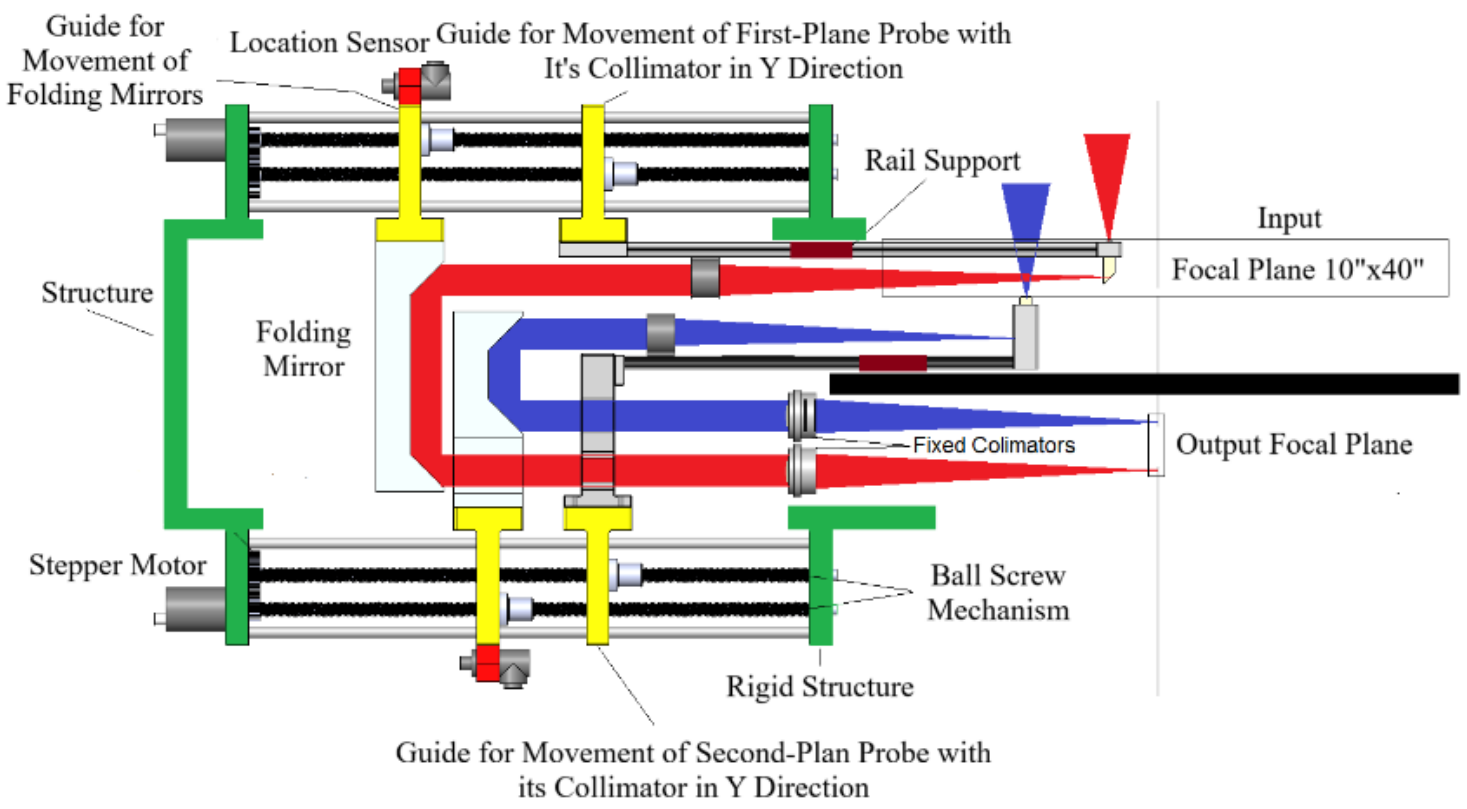

**Fig 5**: Proposed scanning mechanism for the VESPER focal plane.

### A. Main Box

The main box serves as the elements' support, ensuring that all optical and mechanical components are maintained in their designated positions with the required alignment accuracy. As illustrated in Figure 6, the structure is composed of four main sections, two upper and two lower parts, joined at an intermediate interface using precision alignment pins and bolted connections. This configuration facilitates assembly. It is designed as a monolithic structure manufactured from aluminum alloy (series 6 or 7) plates, selected for their favorable strength per weight ratio, machinability, and thermal performance at cryogenic temperatures.

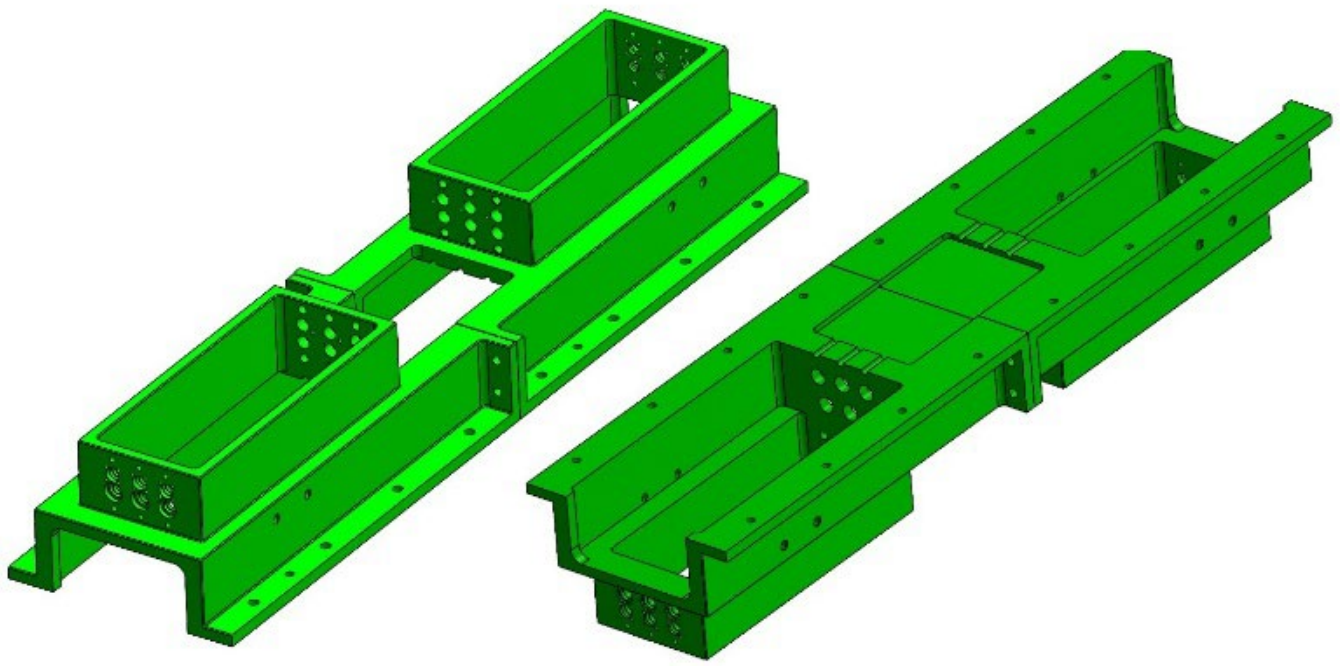

**Fig. 6** Mechanical design of the main structural box, composed of upper and lower sections connected via alignment pins and bolted joints. The structure houses optical elements on two distinct planes while maintaining alignment and stiffness.

The upper section accommodates the prisms and associated components located on the first optical plane, while the lower section supports the prisms and components of the second optical plane. This layered arrangement allows efficient packaging of the optical elements while maintaining clear separation between optical paths and simplifying integration and alignment procedures.

#### B. First plan assembly (Upper Box)

As mentioned, the probes are arranged in two planes, for the first plane, six probes, together with their associated optical, mechanical, and electrical components, are integrated within the upper structural section. These probes are symmetrically distributed, with three units positioned on each side of the structure. The positioning and motion of the probes are controlled by a precision guiding system consisting of linear guide bars and a lead screw mechanism, actuated by cryogenic-compatible stepper motors. In Figure 7 the detail of assembly is shown.

#### C. Second plan assembly (Bottom Box)

An additional set of six probes is arranged on the second optical plane and integrated within the lower structural section (bottom box), as illustrated in Figure 8. The overall design concept is similar to that of the upper assembly; however, differences arise in the configuration of the pick-off probes and the associated folding mirrors, which are adapted to the geometry and optical path of the second plane. This assembly also incorporates the second focal plane, which serves as the collection interface to direct the selected light toward the VESPER spectrograph. The bottom box is therefore a critical element in the optical chain, ensuring proper beam routing, alignment, and compatibility with the downstream image slicing system.

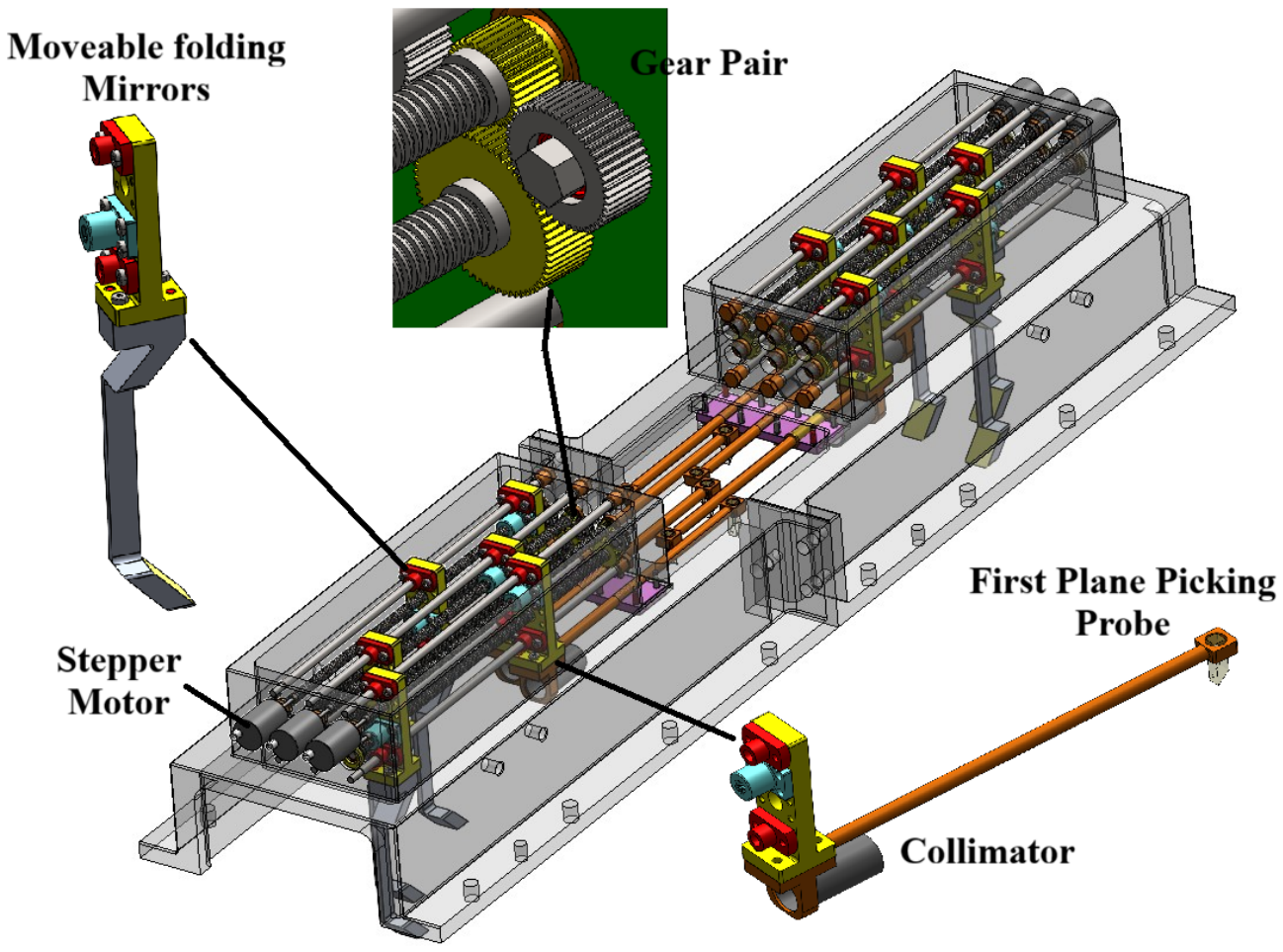

**Fig. 7** Detailed assembly of the probe system, showing the arrangement of the probes in first planes and the associated guiding and actuation mechanisms.

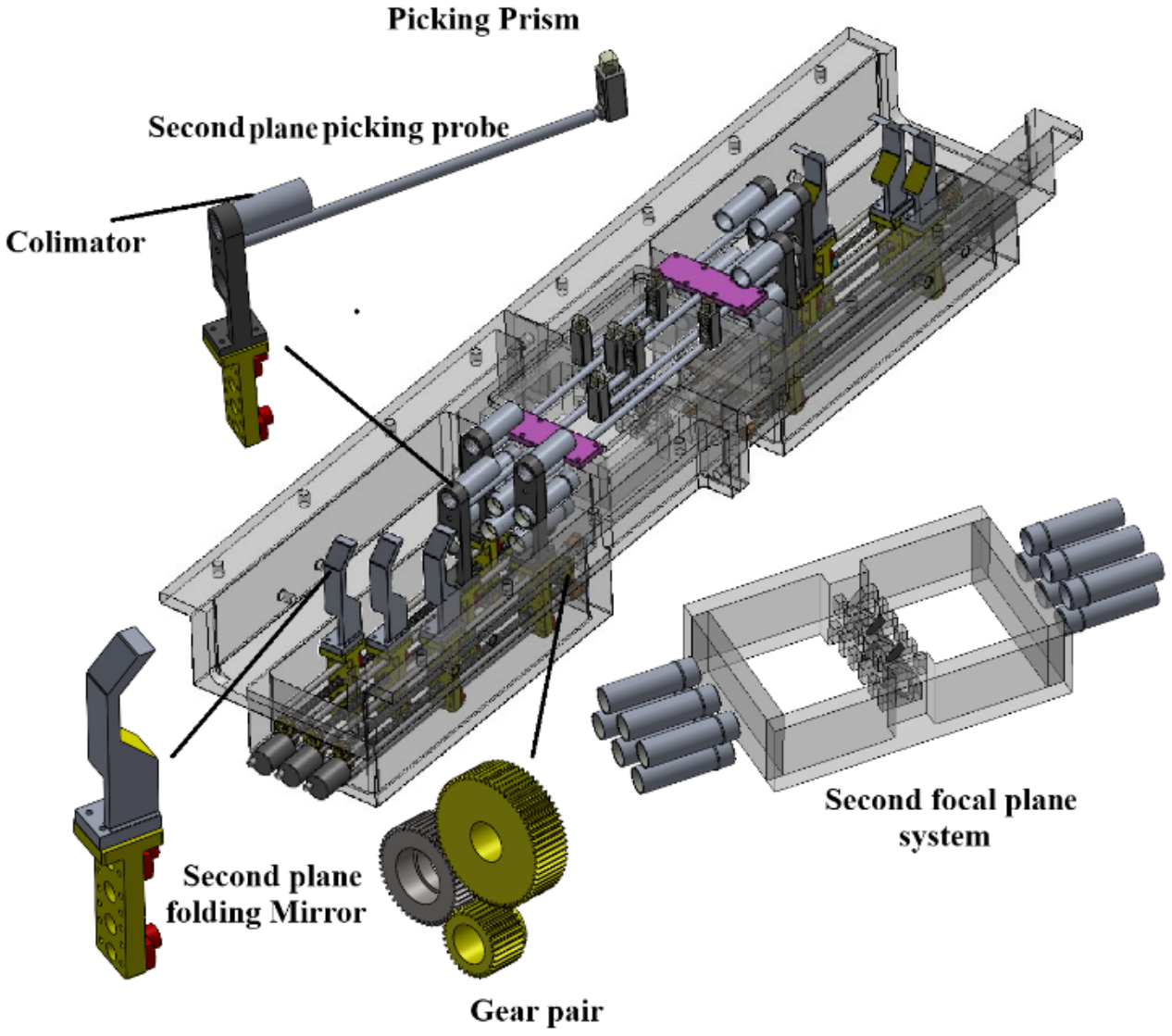

**Fig. 8** Second plane assembly (bottom box), showing the arrangement of the six probes, associated folding mirrors, and the focal plane directing the light toward VESPER.

After completion of the upper and lower assemblies, they are joined through an intermediate flange, forming the complete structure. The components are designed such that each probe can operate independently, without mechanical or optical interference with adjacent probes or other subsystems. A complete model of the assembly, with a cross-sectional view, is presented in Figure 9. As shown, the mechanical layout allows the folding mirrors associated with different probes to pass within close proximity without interference, demonstrating the effectiveness of the compact integration strategy.

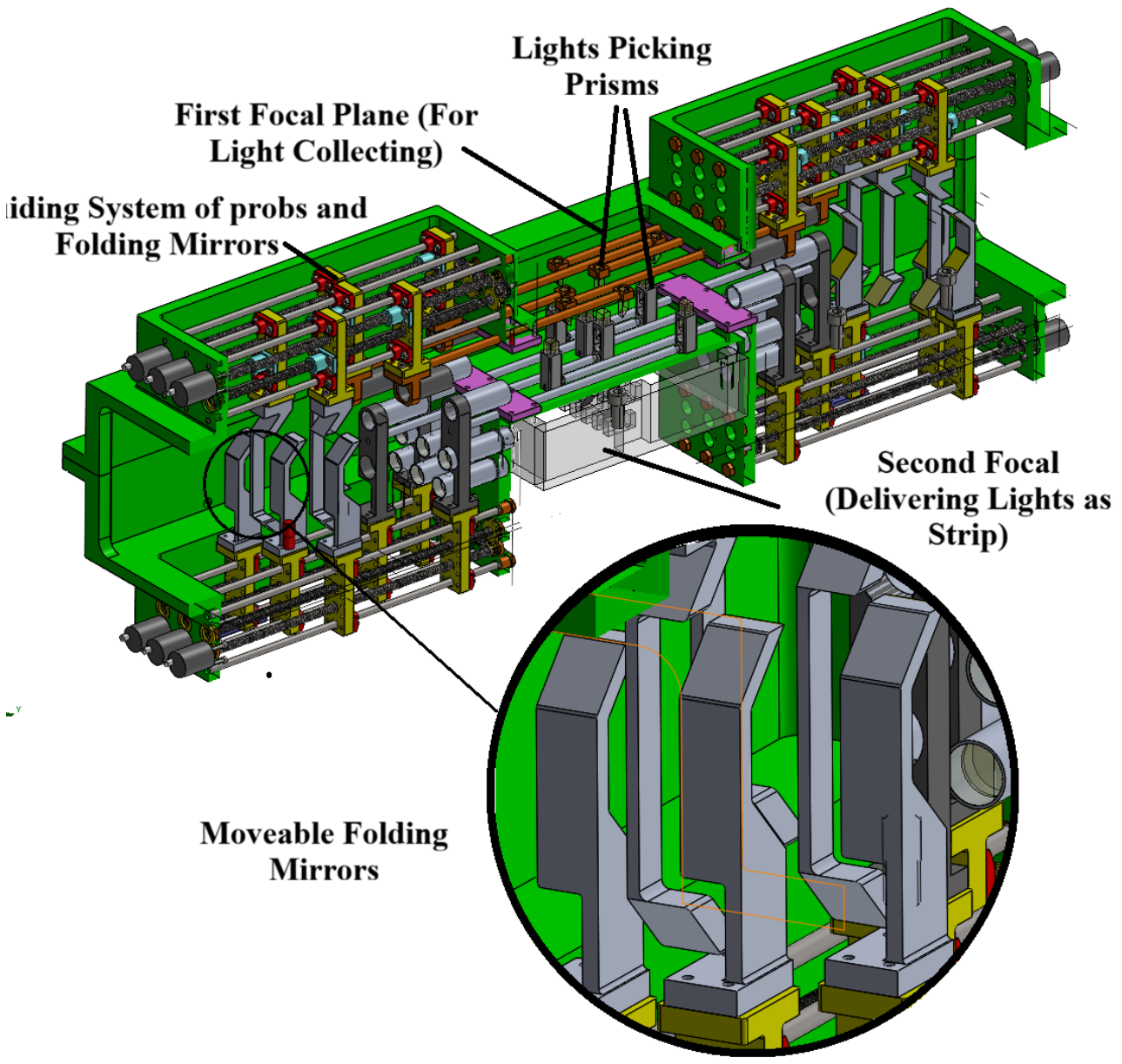

**Fig. 9** Complete assembly of the probe system, showing the integration of the upper and lower boxes connected via the intermediate flange, along with a cross-sectional view highlighting the compact arrangement and non-interfering optical paths.

## 4. PROTOTYPE AND LESSONS LEARNED

To evaluate the system performance and identify potential areas for improvement, a prototype representing one quarter of the system, including a complete probe and its associated components, was manufactured and experimentally tested.

### A. Experimental Setup

The designed experimental setup is shown in Figure 10. In this configuration, the light-collecting prism and its corresponding collimator, as implemented in the final instrument design, were replaced by a fiber-coupled miniature collimator. This collimator, together with the probe, scans the focal plane and emulates the functionality of the light-collecting prism.

At the location corresponding to the second focal plane, where the collected light is delivered in the nominal optical design, a camera was installed. As in the main optical configuration, an achromatic doublet was positioned in front of a camera to produce a point-source image, as illustrated in Figure 11. The stability and positional fluctuations of this point source on the detector as a function of the movements of the FS and the folding mirrors were measured to evaluate the system accuracy and to identify the factors affecting its performance.

The equipment used in the experimental setup is listed in Table 4.

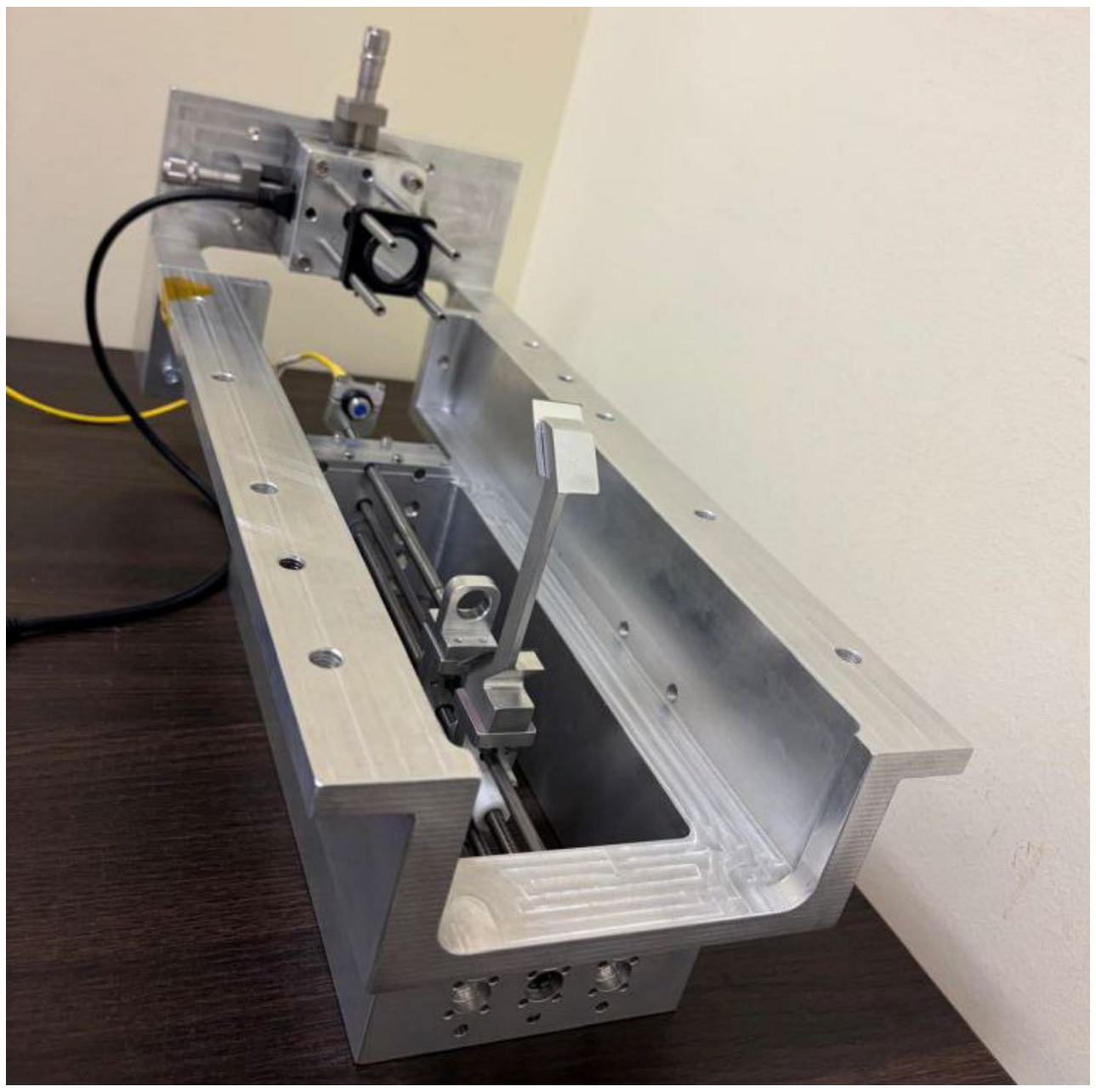

**Fig. 10** Experimental setup developed for the prototype validation. The light-collecting prism of the nominal design is replaced by a fiber-coupled collimator that scans the focal plane together with the probe, while a camera located at the second focal plane records the resulting point-source image for stability and accuracy measurements.

**Table 4: List of equipment and instrumentation used in the prototype experimental setup.**

| | |
|---|---|
| Light collector | Labsphere's integrating sphere |
| Fiber and collimator | F280FC-B with 600-800nm fiber |
| Focusing doublet | THORLABS AC-075-A-ML |
| Adjusting table | - |
| Camera | BASLER da A720-520um 520×720 Pixel (6.9 um Pixel Size) |

## 5.TEST PROCEDURE AND RESULTS

At this stage, the experimental tests were conducted at ambient temperature using commercial off-the-shelf components for the bearings, linear guides, and lead-screw mechanism. The primary objective was to perform an initial evaluation of the system performance and to identify potential issues and areas for improvement.

Cryogenic testing was deferred to future development phases, during which a more refined and optimized version of the system will be implemented. This phased approach enables the validation of the fundamental design concepts before proceeding to testing under operational cryogenic conditions.

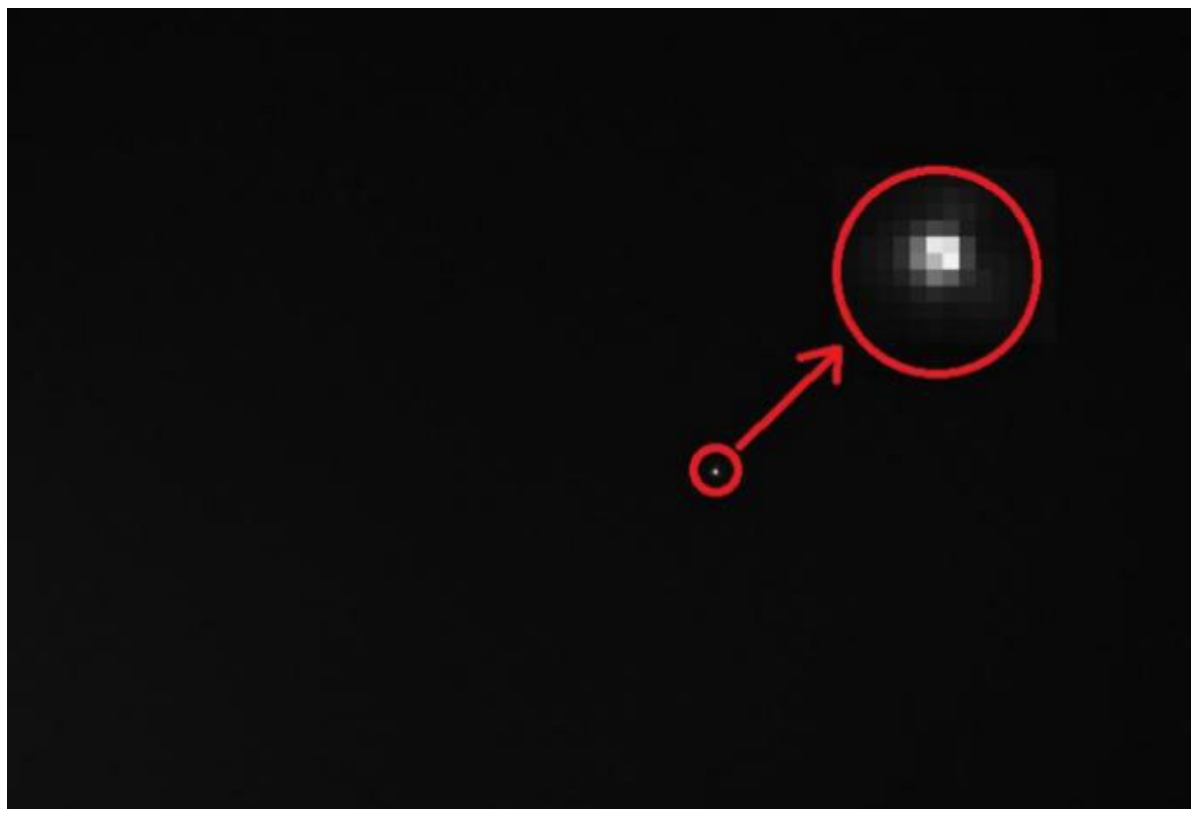

**Fig. 11** Point-source image recorded by the camera detector during the experimental test. The measured spot size is approximately 4 pixels, demonstrating the imaging quality achieved by the optical setup.

### A. Procedure

The experimental data were collected manually. Instead of using the stepper motor, the lead screw was actuated through a manual handle, causing the probe to scan the focal plane from one edge to the other. Each complete revolution of the handle corresponded to a linear displacement of 2 mm of the probe. Measurements were first acquired while scanning from the right side of the focal plane to the left side (forward direction). The procedure was then repeated in the opposite direction, from left to right (backward direction). To evaluate the repeatability of the system, this procedure is repeated 5 times and all data is recorded.

At each measurement position, the centroid coordinates of the point-source image formed on the camera detector were recorded. The resulting data were subsequently analyzed to quantify the positional stability and repeatability of the optical system throughout the scan range.

### B. Results

Based on the experimental results, two main sources of error were identified. The first one is a local tilt of the probe and folding mirrors assembly caused by the mechanical clearance present in the moveable mount and rail system. The second one is a position-dependent tilt resulting from imperfections in the straightness of the rail system, which vary as the probe moves across the scan range. As illustrated in Figure 12, two angular deviations were considered: a vertical tilt, denoted by (θ), associated with the measured displacement (Δz), and a horizontal tilt, denoted by (β), associated with the measured displacement (Δx). The quantities (Δz) and (Δx) were directly measured from the detector images, and the corresponding angular deviations were then determined using the equivalent optical model of the experimental setup developed in Zemax OpticStudio, as illustrated in Figure 13. Based on the geometry of the design, the measured displacement (Δx) is interpreted as the angular tilt (β) of the folding mirror, whereas the measured displacement (Δz) corresponds to the angular tilt (θ) of the probe assembly.

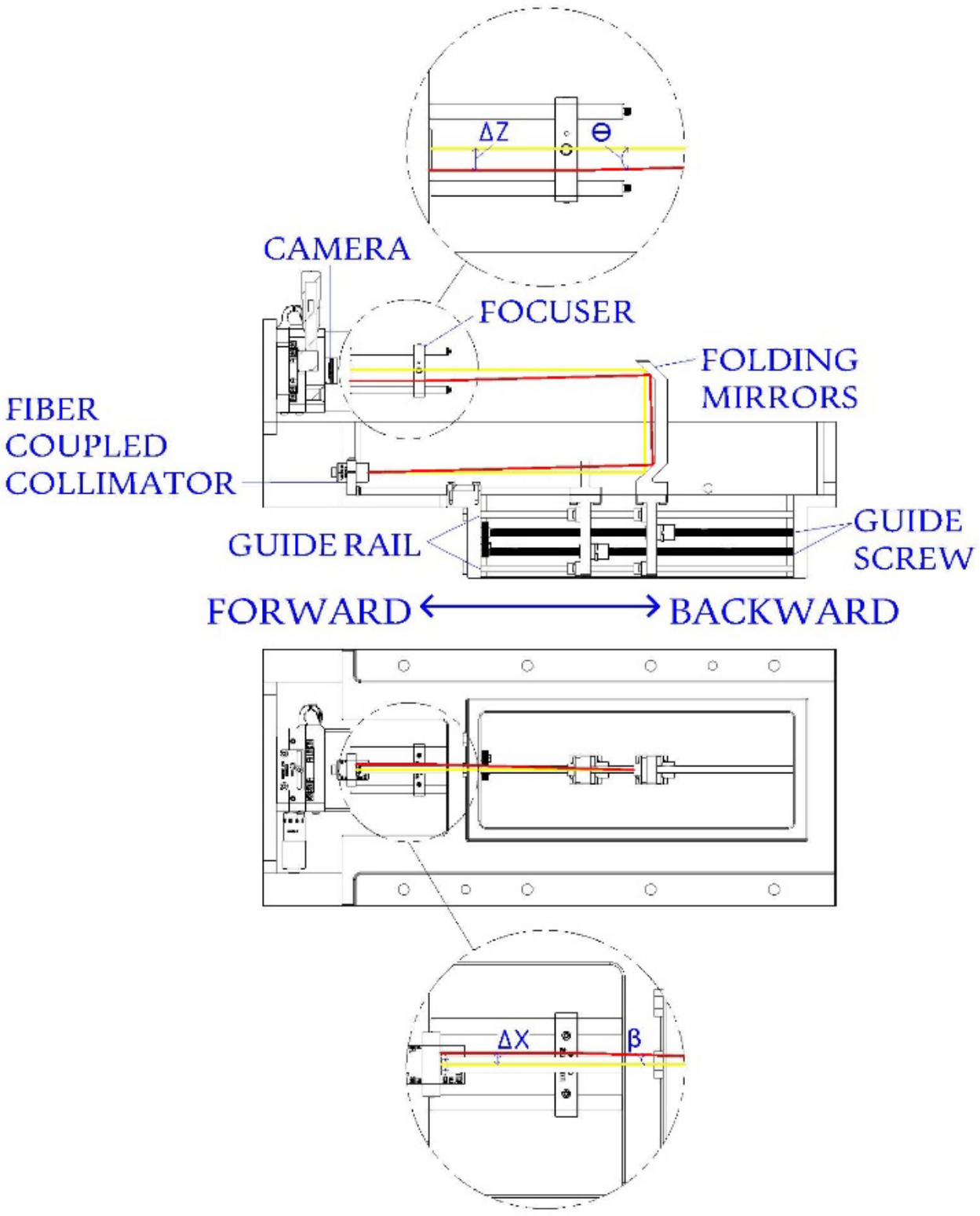

**Fig. 12** Schematic representation of the probe tilt measurement principle. Vertical and horizontal angular deviations, denoted by (θ) and (β), respectively, are derived from the measured image displacements (Δz) and (Δx) on the detector. The equivalent angular errors are calculated from the prototype optical model and are subsequently incorporated into the optical model to evaluate their impact on system performance.

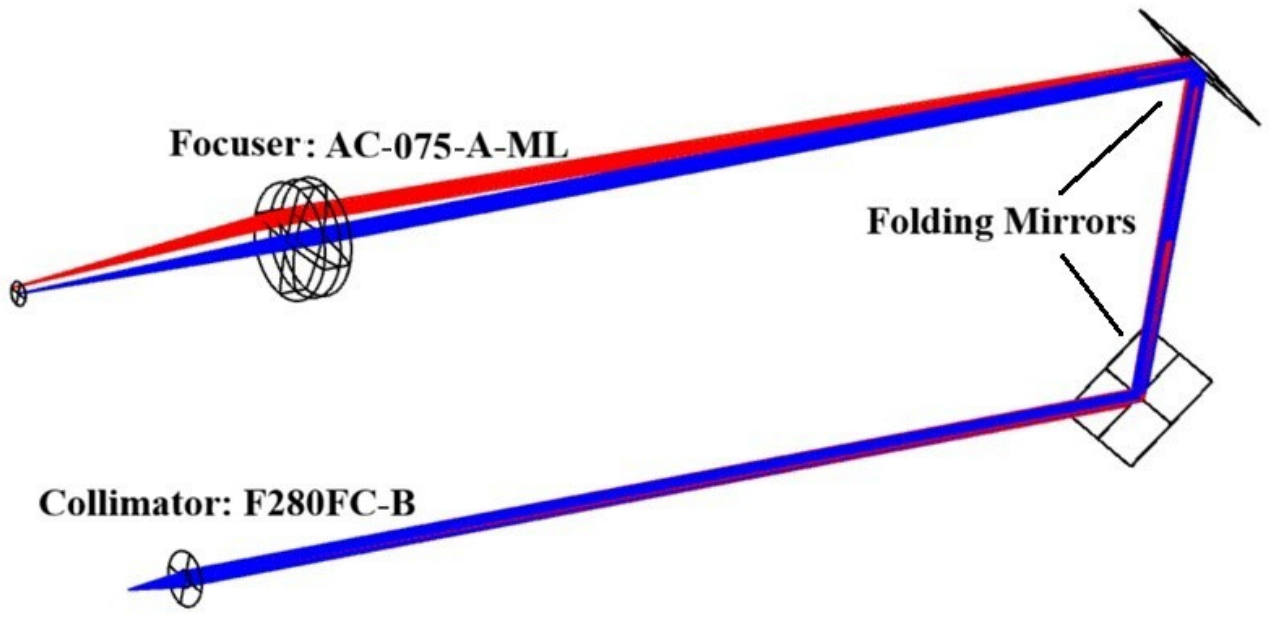

**Fig. 13.** Zemax model of the experimental setup used to determine the equivalent angular deviations from the measured detector image displacements.

The resulting tilt values were then introduced into the VESPER optical model by applying equivalent angular perturbations to the Field Selector System (FSS). This approach enabled the assessment of the impact of the measured mechanical errors on the optical performance of the instrument. More generally, the methodology is not limited to SHARP and can be applied to the tolerance analysis of other Integral Field Spectrograph (IFS) systems to evaluate the effects of probe positioning and alignment errors on image quality and optical stability.

The results obtained from the full focal-plane scan in both the forward and backward directions are presented in Figure 14. The measurements reveal a significant local tilt in the x-direction (Δx), which is primarily attributed to the mechanical clearance in the moveable parts and rail system. The maximum measured displacement fluctuates within approximately ±0.16 mm, corresponding to an equivalent angular deviation (tilting of folding mirrors) of ±0.061° during the forward scan, and ±0.1 mm corresponding to ±0.038° during the backward scan. The position-dependent tilt associated with the probe motion across the focal plane, which is mainly caused by straightness errors in the rail system, reaches a maximum displacement of ±0.23 mm, equivalent to an angular deviation of ±0.088°.

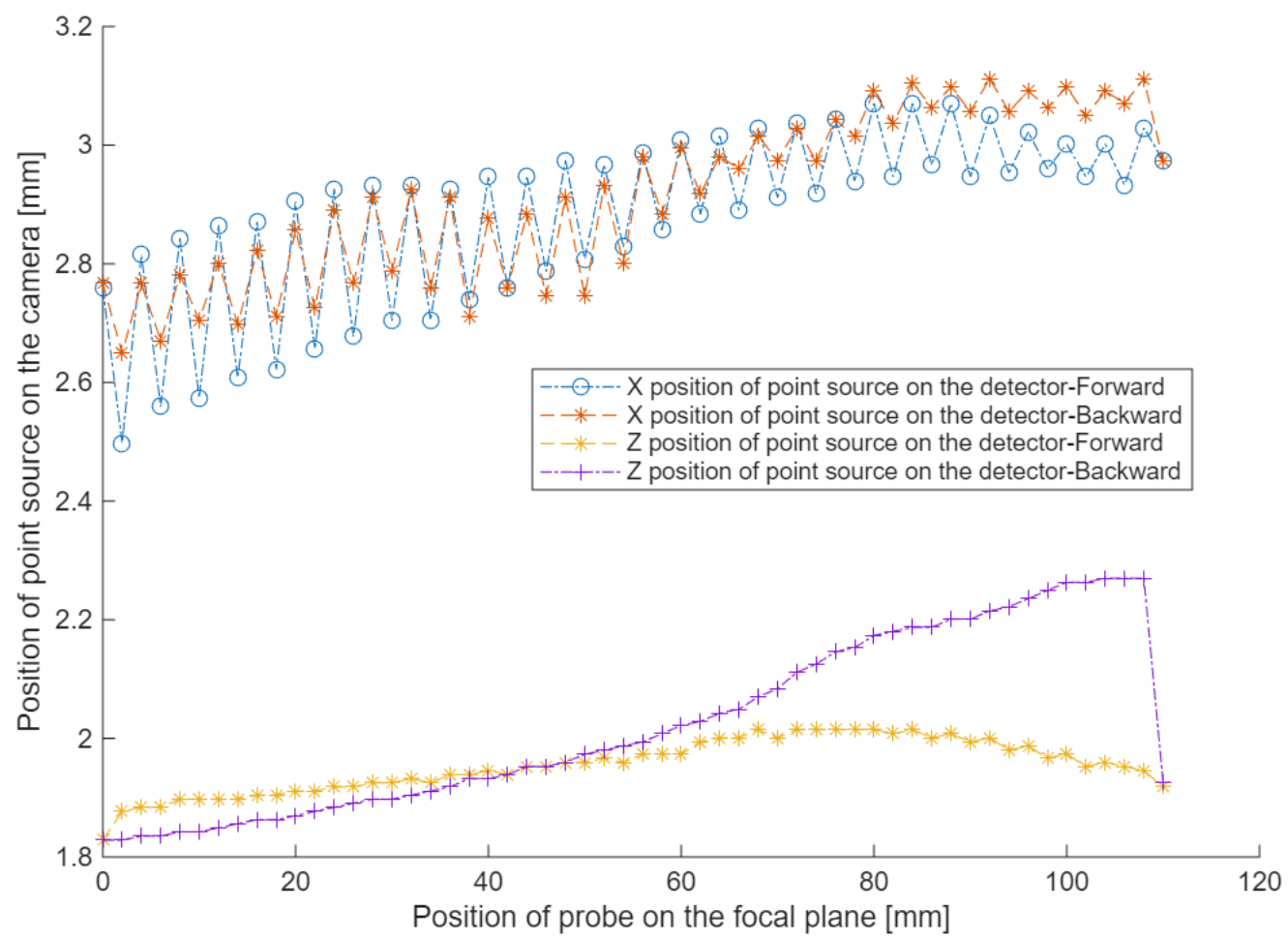

**Fig. 14** Measured point-source displacements during a full scan of the focal plane in the forward and backward directions. The resulting detector shifts (Δz) and (Δx)) are used to derive the equivalent angular deviations of the probe assembly.

**Table 5: Summary of the measured displacement and equivalent angular deviations for the identified error sources, together with their corresponding impact on the optical performance of the SHARP instrument.**

| Parameter | Forward | Backward | Comment |
|---|---|---|---|
| **Maximum local Δx [mm]** | ±0.16 | ±0.10 | — |
| **Maximum local Δz [mm]** | <±0.01 | <±0.01 | — |
| **Maximum local β [°]** | ±0.061 | ±0.038 | Mirror stability |
| **Maximum local θ [°]** | <±0.008 | <±0.008 | Fiber–collimator stability |
| **Maximum positioning Δx [mm]** | ±0.23 | ±0.17 | Due to the rail and structure |
| **Maximum positioning Δz [mm]** | ±0.22 | ±0.22 | Due to the rail and structure |
| **Maximum positioning β [°]** | ±0.088 | ±0.065 | Mirror tilt due to positioning |
| **Maximum positioning θ [°]** | ±0.169 | ±0.169 | Fiber–collimator positioning tilt |
| **Total Δx [mm]** | ±0.30 | — | Combined positioning/local error |
| **Total Δz [mm]** | ±0.22 | — | Combined positioning/local error |
| **Total β [°]** | ±0.115 | | Combined mirror tilts and fiber–collimator tilt |
| **Total θ [°]** | ±0.169<br>— | | Combined mirror tilts and fiber–collimator tilt |
| **Max. Output angular deviation in FFS β [°]** | ±0.46* | | |
| **Max. Output angular deviation θ [°]** | ±1.05* | | |
| **Effect on VESPER** | The rough values of the spot radius increase (on the detector) due to the maximum tilt allowable is: +0.2 um with 0.05° of θout tilt, +0.4 um with 0.01° of βout tilt | | |

* These values are related to the prototype which are roughly 3 times faster that the real FSS, therefore in the real FSS the max angular deviations will be much smaller:

In contrast, the tilt in the z-direction (Δz) remains negligible throughout most of the scanning range. A noticeable deviation is observed only at the reversal point on the left side of the focal plane, where the scanning direction changes (fiber+collimator tilting). At this location, the maximum displacement reaches ±0. 22 mm, corresponding to an equivalent tilt of ±0.169°.

These results indicate that further improvement of the system is required to meet the specified optical requirements. The measured displacements and equivalent angular deviations for the different error scenarios, together with their impact on the output beam and the corresponding VESPER limitation, are summarized in Table 5.

To quantitatively evaluate the repeatability of the scanning mechanism, the experiment was repeated five times in both the forward and backward directions. The resulting displacement profiles in the x-direction for the forward scans and in the z-direction for the backward scans are presented in Fig. 15. At each probe position, the mean value and standard deviation of the five measurements were calculated. The resulting statistics are summarized in Table 6. For the forward x-direction measurements, the mean standard deviation over the scanning range was 0.0178 mm, with a maximum value of 0.0410 mm. For the backward z-direction measurements, the corresponding mean and maximum standard deviations were 0.0070 mm and 0.0215 mm, respectively. The maximum peak-to-peak variations were 0.1035 mm and 0.0552 mm for the forward x and backward z-measurements, respectively. These results provide a quantitative assessment of the repeatability of the scanning mechanism under the tested ambient-temperature conditions.

**Table 6: Quantitative repeatability statistics obtained from five repeated scans in the forward and backward directions. (F=forward, B=backward)**

| Statistic | X–F | X–B | Z–F | Z–B |
|---|---|---|---|---|
| Mean SD [mm] | 0.0178 | 0.0233 | 0.0049 | 0.0070 |
| Maximum SD [mm] | 0.0410 | 0.0574 | 0.0234 | 0.0215 |
| Maximum peak-to-peak [mm] | 0.1035 | 0.1449 | 0.0552 | 0.0552 |
| Maximum deviation [mm] | 0.0718 | 0.1007 | 0.0414 | 0.0345 |

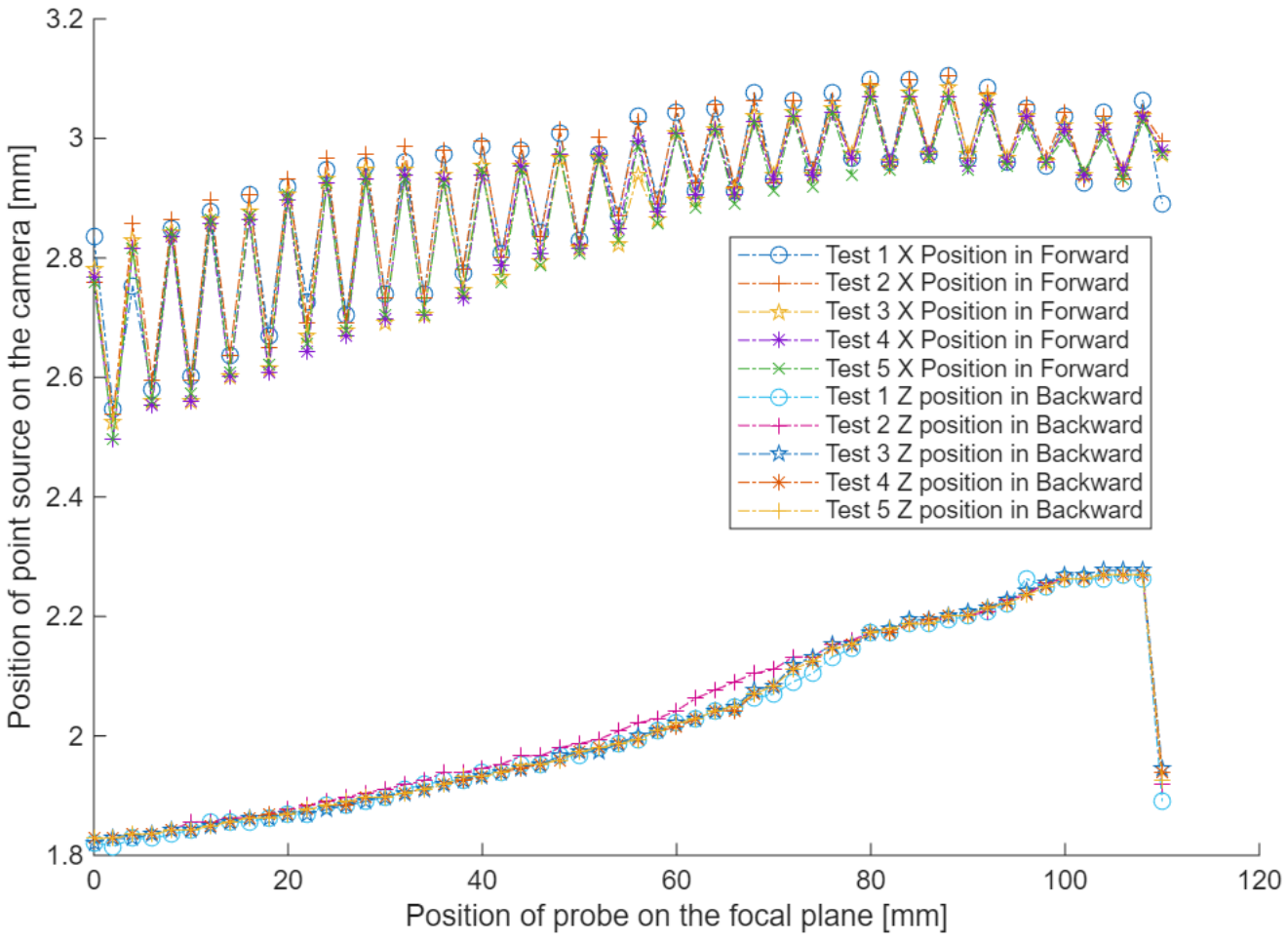

**Fig. 15** . Measured detector displacement as a function of probe position for five repeated scans in the forward and backward directions. The overlap between the measurement curves demonstrates the repeatability of the probe motion and the consistency of the mechanical system.

**C. Potential improvement**

Based on the experimental results, the most significant local tilt was observed in the (x)-direction (Δx), as illustrated in Figure 12 (top view). As shown in the figure, the two linear rails are located on the same plane, making the system more susceptible to rotational motion about axis perpendicular to movement direction. Consequently, the use of higher-precision rectangular linear guides, together with an increased separation between the rails in the (y)-direction, is expected to significantly reduce this source of error.

The local tilt in the z-direction becomes particularly pronounced when the probe reaches the end of its travel range at the left edge of the focal plane (end of forward direction). As illustrated in the front view of Figure 12, the probe is effectively supported just from only one side at this position, allowing the assembly to rotate about the support point. This effect can be mitigated by extending the probe arm in the opposite direction and introducing an additional support point. Furthermore, increasing the spacing between the linear guides would improve the rotational stiffness of the mechanism and further reduce the observed tilt.

In addition, the smooth transmission of force through the lead-screw mechanism plays a critical role in ensuring stable motion of the movable assembly. Any irregularities in force transmission, such as backlash, friction variations, or misalignment within the drive system, can introduce unwanted perturbations that contribute to probe tilting and positioning errors. Therefore, optimizing the lead-screw mechanism and minimizing mechanical clearance can improve the smoothness of the probe motion and reduce angular deviations in both the horizontal and vertical directions.

The position-dependent tilt associated with the probe motion is primarily influenced by manufacturing and assembly errors of the supporting structure. This error can be minimized by improving the straightness of the guide system and by adopting alignment-based assembly procedures. In such an approach, the final rail positions are established through precision alignment during assembly rather than relying solely on the manufacturing accuracy of the supporting structure. This methodology enables compensation for fabrication tolerances and can substantially improve the overall positioning accuracy and stability of the scanning mechanism. Since the final FSS is intended to operate under cryogenic conditions, these effects are expected to become more critical due to thermal contraction, changes in guiding clearances, and differential deformation between the rails and the supporting structure. Therefore, in the next phase of development, the tolerances between the guiding rails and the corresponding guiding components will be defined by considering the expected cryogenic dimensional changes and the resulting operating conditions. As an alternative solution, ceramic bushing systems with low coefficients of thermal expansion will be investigated to minimize variations in the guiding clearances over the operating temperature range. In addition, the connection between the guiding rails and the supporting structure will be redesigned using compliant/flexible interfaces to reduce the transmission of structural deformation and differential thermal contraction to the guiding system.

## 3. Conclusion

This paper presented the design, prototyping, and experimental evaluation of a novel field-selection system developed for the VESPER channel of the SHARP near-infrared multi-mode spectrograph. The proposed concept employs movable light-collection probes to select targets across the telescope focal plane while maintaining optical performance and operational flexibility. The optical and mechanical architecture of the system was described, and a prototype representing one quarter of the final assembly was manufactured and tested to evaluate the feasibility of the proposed approach. A dedicated experimental setup was developed to characterize the positioning stability and repeatability of the probe motion. The measurements revealed two primary sources of error: local probe tilting caused by mechanical clearances in the movable mount and guide system, and position-dependent tilting resulting from straightness errors in the rail assembly. The measured image displacements were converted into equivalent angular deviations and compared with the maximum allowed values preliminarily defined for VESPER, to assess their impact on instrument performance. The experimental results demonstrated good repeatability of the scanning mechanism and provided valuable insight into the relationship between mechanical tolerances and optical performance. The analysis identified several opportunities for improvement, including the use of higher-precision linear guides, increased rail separation, improved probe support, optimized lead-screw force transmission, and alignment-based assembly procedures. These modifications are expected to significantly reduce positioning and tilting errors and enhance system stability.

Although the present study was conducted at ambient temperature using commercial off-the-shelf mechanical components and not fulfill the requirement of the VESPER, the results provide a solid foundation for the next development phase and demonstrate the preliminary feasibility of the proposed FSS concept at the conceptual-design and ambient-test level, but do not constitute a complete validation of its performance under cryogenic operating conditions. The effects of differential thermal contraction, temperature-dependent mechanical behavior, and the suitability of the components for cryogenic operation will be addressed during the detailed design and subsequent prototype development. Cryogenic experimental validation of the optimized system will constitute the next stage of development. The future work will focus on implementing the proposed design improvements, performing cryogenic validation tests, and evaluating the system under operational conditions representative of the final instrument. Overall, the results provide a basis for further development of the proposed field-selection concept for SHARP and demonstrate its potential applicability to other multi-IFU spectrographs requiring precise focal-plane target selection.

## Acknowledgments

HM, PS and the SHARP team acknowledge the support from the INAF Techno-Grant 2022, SHARP - 1.05.12.02.01; and the INAF Large-Grant 2024, SHARP - 1.05.24.01.01.

## Disclosures

The authors declare no conflicts of interest.

## Data availability

All data in support of the findings of this paper are available within the article. More information can be found on this website: https://sharp.brera.inaf.it/.